\documentclass[12pt]{article}
\usepackage{a4,amsmath,amssymb,cite,graphicx,multirow,cancel}
\def\Bbb{\mathbb}
\def\BZ{\Bbb Z} \def\BR{\Bbb R}

\def\Tr{\mathrm{Tr}}

\newcommand{\etabox}[2]{\underset{\ ~#2}{\mbox{\scriptsize $#1$}~\framebox[15pt]{\phantom{a}}}}

\newcommand{\wt}[1]{\widetilde{#1}}
\catcode`@=11 \@addtoreset{equation}{section} \catcode`@=12

\begin{document}
\bibliographystyle{utphys}
\begin{titlepage}
\renewcommand{\thefootnote}{\fnsymbol{footnote}}
\noindent
{\tt IITM/PH/TH/2011/3}\hfill
{\tt arXiv:1106.5715 [hep-th]} \\[4pt]
\mbox{}\hfill 
\hfill{\fbox{\textbf{v2.2; June 2018 }}}

\begin{center}
\large{\sf  Unravelling Mathieu moonshine}
\end{center} 
\bigskip 
\begin{center}
{\sf Suresh Govindarajan}\footnote{\texttt{Email: suresh@physics.iitm.ac.in}} \\[3pt]
\textit{Department of Physics, Indian Institute of Technology Madras,\\ Chennai 600036, India \\[4pt]
}
\end{center}
\bigskip
\bigskip
\begin{abstract}
The $D1-D5-KK-p$ system naturally provides an infinite dimensional module graded by the dyonic charges whose dimensions are counted by the Igusa cusp form, $\Phi_{10}(\mathbf{Z})$. We show that the Mathieu group, $M_{24}$, acts on this  module by recovering the Siegel modular forms that count twisted dyons  as a trace over this module. This is done by recovering Borcherds product formulae for these modular forms using the $M_{24}$ action. This establishes the correspondence (`moonshine') proposed in arXiv:0907.1410 that relates conjugacy classes of $M_{24}$  to Siegel modular forms. This also, in a sense that we make precise, subsumes existing moonshines for $M_{24}$ that relates its conjugacy classes to eta-products and Jacobi forms. 

\end{abstract}
\texttt{June 2018: Corrected  some sign errors that appear in the published version as well.}
\end{titlepage}
\setcounter{footnote}{0}

\section{Introduction}

In ref. \cite{Govindarajan:2009qt}, a moonshine for the Mathieu group, $M_{24}$, relating its conjugacy classes to genus-two Siegel modular forms was proposed.  Let $\rho=1^{a_1}2^{a_2}\cdots N^{a_N}$ be the cycle shape for an $M_{24}$ conjugacy class and $(k+2)=\tfrac12 \sum_i a_i$. Then the moonshine correspondence proposed in \cite{Govindarajan:2009qt} (see also \cite{Govindarajan:2010fu}) is as follows:
\begin{equation}
\rho \longrightarrow \Phi^\rho_k(\mathbf{Z})\ ,
\end{equation}
where $ \Phi^\rho_k(\mathbf{Z})$ is a genus two Siegel modular form at level $N$ and $\mathbf{Z} = \left(\begin{smallmatrix} \tau & z \\ z & \sigma \end{smallmatrix}\right)\in \mathbb{H}_2$. We shall restrict the considerations of this paper to the situation when the conjugacy class reduces to a conjugacy class of $M_{23}$: this implies that $a_1\neq 0$. In these cases, it is known that the Siegel modular form counts $\tfrac14$-BPS  twisted dyons in the heterotic string compactified on a six-torus.\footnote{Such dyons are invariant under the action of a finite abelian group of order $N$ -- this group maps to a symplectic automorphism of $K3$ in the dual type II picture.}  
 
It is natural to ask whether there exists an $M_{24}$-module $V^\natural$ graded by three integers $(n,\ell,m)$ corresponding to the dyonic charges $(\tfrac12\mathbf{q}_e^2,\mathbf{q}_e\cdot \mathbf{q}_m,\tfrac12\mathbf{q}_m^2)$ i.e.,
\begin{equation}
V^\natural = \bigoplus_{(n,\ell,m)} V_{n,\ell,m} \ ,
\end{equation}
such that the Igusa cusp form is given by the following trace\footnote{The superdimension sdim of the graded vector space $V=V_+\oplus  V_-$ is defined to be $\textrm{sdim}(V)=\textrm{Tr}_V(-1)^F:=\textrm{STr}_V(1)=(d_+-d_-)$ where $d_+$ is the dimension of the bosonic (even) subspace, $V_+$ and $d_-$ is the dimension of the fermionic (odd) subspace, $V_-$ of $V$.}
\begin{equation}
\frac1{\Phi_{10}(\mathbf{Z})} = \sum_{(n,\ell,m)}~ \textrm{sdim}\left[V_{(n,\ell,m)}\right] q^n r^\ell s^m \ ,
\end{equation}
with $q=\exp(2\pi i\tau)$, $r=\exp(2\pi iz)$, $s=\exp(2\pi i \sigma)$.
Further,
\begin{equation}
\frac1{\Phi^\rho_{k}(\mathbf{Z})} = \sum_{(n,\ell,m)}~ \textrm{STr}\left[g\big|_{V_{(n,\ell,m)}}\right] q^n r^\ell s^m \ ,
\end{equation}
counts $g$-twisted $\tfrac14$-BPS states in the toroidally compactified heterotic string and $\rho$ is the conjugacy class of $g\in M_{24}$ that reduces to a conjugacy class of $M_{23}$.

The $D1-D5-KK-p$ system\cite{David:2006ji} provides an obvious candidate for $V^\natural$. It has a natural decomposition into three distinct parts:
\begin{equation}
V^\natural = W \otimes \mathcal{B} \otimes \mathcal{S} \ ,
\end{equation}
with
\begin{align*}
\sum_{(n,\ell)}~ \textrm{sdim}\left[W_{(n,\ell,0)}\right] q^n r^\ell s^0&=  \frac{\eta(\tau)^6}{\theta_1(\tau,z)^2}\ ,\\
\sum_{n}~ \textrm{sdim}\left[\mathcal{B}_{(n,0,0)}\right] q^n r^0 s^0 &= \frac1{\eta(\tau)^{24}} \ ,\\
\sum_{(n,\ell,m)}~ \textrm{sdim}\left[\mathcal{S}_{(n,\ell,m)}\right] q^n r^\ell s^m&=   \frac1{\mathcal{E}(K3,\mathbf{Z})}\ ,
\end{align*}
where  $\mathcal{E}(K3,\mathbf{Z})$ is the second-quantized elliptic genus of $K3$\cite{Dijkgraaf:1996xw}. The problem thus reduces to showing the action of $M_{24}$ on these three modules.

The paper is organized as follows. In section 2, we discuss the module $V^\natural$ in terms of its components. When possible, we show how some parts are $M_{24}$-modules connecting them to known moonshines for $M_{24}$. We reduce the problem to proving that the module $\mathcal{S}$, associated with  the second-quantized elliptic genus of $K3$, is an $M_{24}$ module. In section 3, we focus on the details of the moonshine proposed in \cite{Govindarajan:2009qt}.  We show that the module $\mathcal{S}$ is an $M_{24}$ module  by showing that the $M_{24}$ action implies replication formulae for twisted elliptic genera of symmetric products of $K3$. It is proposed that a twisted Hecke-like operator generates all these replication formulae. In section 4, we prove that the proposal provides a Borcherds product formula that incorporates the $M_{24}$-action on the second-quantized elliptic genus of $K3$ and hence for the Siegel modular forms. This product formula agrees with existing formulae in the literature. We conclude with a few observations. Appendix A provides some information about the elliptic genus of K\"ahler and hyperK\"ahler manifolds.

\section{The module $V^\natural$ and its parts}

The microscopic counting of $\tfrac14$-BPS states in four-dimensional CHL compactifications of the heterotic string with $\mathcal{N}=4$ supersymmetry was carried out by David and Sen\cite{David:2006yn}. Using a chain of dualities as well the 4D-5D correspondence\cite{Shih:2005uc}, the counting was carried out in a configuration of  D1 and D5 branes moving in $K3\times S^1\times TN$ where $TN$ is the Taub-NUT space. The geometry  close to the center of $TN$ is five-dimensional while the geometry far away from the center is four-dimensional with the appearance of an additional circle $\widetilde{S}^1$.

The result of David-Sen\cite{David:2006yn} for the type IIB dual of the toroidally compactified heterotic string expresses  the Igusa cusp form, $\Phi_{10}(\mathbf{Z}),$ as a product of three terms
\begin{equation}
\frac{64}{\Phi_{10}(\mathbf{Z})} = \underset{\textrm{(i)}}{\underbrace{\bigg[\frac{4\ \eta(\tau)^6}{\theta_1(\tau,z)^2}\bigg]}} \times 
\underset{\textrm{(ii)}}{\underbrace{\bigg[\frac{16}{\eta(\tau)^{24}}\bigg]}}
\times \underset{\textrm{(iii)}}{\underbrace{\bigg[\frac1{\mathcal{E}\!\left(K3;\mathbf{Z}\right)}\bigg]}}\ ,
\end{equation}
As indicated above the  states that contribute to the counting arise from three distinct sectors (labeled (i)-(iii)) in the type IIB description: 
\begin{itemize}
\item[(i)] the overall motion of  the $D1$-$D5$ branes in Taub-NUT space.
\item[(ii)] the excitations of the KK-monopole -- following the chain of dualities, these excitations get mapped to the states of the heterotic string. 
\item[(iii)] the motion of the $D1$-branes in the worldvolume of the $D5$-branes -- this counting leads to  the second-quantized elliptic genus of $K3$. 
\end{itemize}
In terms of fermionic zero-modes, half of the supersymmetry (contributing sixteen zero-modes) is broken by the KK-monopole and its excitations. The remaining quarter of the supersymmetry is broken (contributing four zero-modes) by the center of mass motion. The motion of the $D1$-branes in the worldvolume of the $D5$-branes do not break any further supersymmetry. Let $Q_1$ and $Q_5$ denote the number of $D1$-branes and $D5$-branes respectively. Further, let $n$ denote the momentum along $S^1$ (the KK monopole charge) and $J$ the angular momentum in $TN$ space. Then, the $D1-D5-KK-p$ system carries the following T-duality invariant charges\cite{David:2006yn}:
\begin{equation}
\tfrac12\mathbf{q}_e^2 =n \ ,\quad \mathbf{q}_e\cdot \mathbf{q}_m=J\ ,\quad
\tfrac12\mathbf{q}_m^2=(Q_1-Q_5)Q_5\ .
\end{equation}
The module $V^\natural$ is graded by these three charges.

We shall now consider each one of these terms separately to further elucidate the structure of the module $V^\natural$. 
The rest of this section is based on ref. \cite{David:2006yn} and we refer the reader to it for a detailed description.

\subsection{The overall  motion in Taub-NUT space}

The overall motion of the $D1$-$D5$ system in Taub-NUT space is described by a 1+1-dimensional supersymmetric conformal field theory(CFT) with Taub-NUT as its target space. The charge associated with $\mathbf{q}_e^2$ gets related to the $L_0$ eigenvalue while the charge associated with $\mathbf{q}_e\cdot \mathbf{q}_m$ gets related to the charge due to a $U(1)_L$ symmetry in the field theory. The field theory consists of four free left-moving fermions; four right-moving fermions charged under $U(1)_L$ and interacting with four scalars with Taub-NUT target space. The unbroken supersymmetry acts on the right-movers. Thus BPS states are obtained when the right-movers are in the ground state. The only permitted excitations are from the four left-moving fermions as well as the left-movers of the four scalars.  The index of the supersymmetric states in the field theory which we interpret as a supertrace over $W$ is given by
\begin{equation}
\sum_{(n,\ell)}~ \textrm{sdim}\left[W_{(n,\ell,0)}\right] q^n r^\ell s^0
= Z_{\textrm{free}} (\tau)\ Z_{\textrm{osc}}(\tau,z)\ Z_{\textrm{zero-mode}}(z)\ ,
\end{equation}
where  $Z_{\textrm{free}} (\tau)$ is the contribution of the free left-moving fermions (which are $z$-independent due to their $U(1)_L$ invariance); $Z_{\textrm{osc}}(\tau,z)$ is the contribution from the left-moving oscillators of the four scalars and 
$Z_{\textrm{zero-mode}}(z)$ is the contribution from the bosonic zero-modes (which are $\tau$ independent). Thus, the space $W$ is the direct product of three-spaces and the contributions are:
\begin{align}
Z_{\textrm{free}} (\tau) &= 4 \prod_{n=1}^\infty (1-q^n)^4\ , \\
Z_{\textrm{osc}}(\tau,z) &= \prod_{n=1}^\infty (1-r q^n )^{-2} (1-r^{-1}q^n)^{-2} \ , \\
Z_{\textrm{zero-mode}}(z) &= \frac{r}{(r-1)^2}\ .
\end{align}

\subsubsection*{The action of $M_{24}$ on $W$}

Let $g$ denote a supersymmetry preserving symmetry of the conformal field theory associated with the sigma model with $K3$ target space of finite order $N$. Symplectic automorphisms of $K3$ furnish a large (but not all) class of such examples. A classification of such symmetries has appeared recently where it has been shown that such symmetries  are subgroups of the Conway group, $Co_1$\cite{Gaberdiel:2011fg}. Since $M_{24}$ is a sub-group of $Co_1$, this includes all situations that implicitly or explicitly  appear in this paper. It follows any such $g\in Co_1$ acts \textit{trivially} on the space $W$. This is easy to understand since $W$ arises from the dynamics of motion in the Taub-NUT space and does not `see' the $K3$. We identify symplectic automorphisms of $K3$ with elements of $M_{24}$.  Thus we conclude  that $M_{24}$ has no action on $W$. Thus,
\begin{equation}
\boxed{
 \sum_{(n,\ell,m)}~ \textrm{STr}\left[g\big|_{W_{(n,\ell,m)}}\right] q^n r^\ell s^m =
 \sum_{(n,\ell,m)}~ \textrm{sdim}\left[W_{(n,\ell,m)}\right] q^n r^\ell s^m \ .
 }
\end{equation}

\noindent \textbf{Remarks:} 
\begin{enumerate}
\item 
$Z_{\textrm{zero-mode}}(z)$ has different Fourier expansions depending on whether $|r|<1$ and $|r|>1$:
\begin{equation}
Z_{\textrm{zero-mode}}(z) = \begin{cases}  r+2r^2 + 3 r^3 +\cdots & \textrm{ for }|r|<1\ , \\
r^{-1}+2r^{-2}+3r^{-3}+\cdots & \textrm{ for } |r|>1\ .
\end{cases}
\end{equation}
Physically, this is related to the existence of a wall  of marginal stability at $r=1$, where there is a jump in dyon degeneracy due to additional contributions from two-centered black holes\cite{Sen:2007vb,Dabholkar:2007vk}. A mathematical interpretation can be given by using the connection of the square-root of the Igusa cusp form, $\Delta_5(\mathbf{Z})$, with a BKM Lie superalgebra\cite{Nikulin:1995}. The wall of marginal stability in the physical setting gets mapped to the wall of a Weyl chamber\cite{Cheng:2008fc}.
\item The module $V^\natural$ is thus  dependent on the Weyl chamber. As in the case of finite Lie algebras, the  Weyl chamber is characterized by the choice of Weyl vector.
\end{enumerate}

\subsection{The KK monopole}

The low-energy dynamics of the D1-D5 system localized at the center of the Taub-NUT space can carry momentum along $S^1$ but not along $\widetilde{S}^1$.\footnote{Recall that the geometry of the Taub-NUT space near the centre is that of $\BR^4$ while it is $\BR^3\times \widetilde{S}^1$ far away from the center. Thus the geometry of spacetime is $K3\times S^1\times \BR^4$.} The momentum along $S^1$ contributes to the electric charge $\tfrac12 \mathbf{q}_e^2$ with other charges vanishing. Following the chain of dualities, BPS states of this system get mapped to BPS states of the fundamental heterotic string. The BPS condition requires the supersymmetric right-movers of the heterotic string to be in the ground state and the various electric charges take values in the Narain lattice for heterotic string compactified on a six-torus. In the light-cone gauge, the space of BPS states is that of the oscillator modes of 24 left-moving  chiral bosons. We will denote this space by $\mathcal{B}$. 

The level matching condition relates the electric charge to the level, $n$, of the oscillator excitations\cite{Sen:2005ch}. One has
\begin{equation}
\tfrac12 \mathbf{q}_e^2 = n-1\ , \quad n=0,1,2,\cdots\ .
\end{equation}
One has
\begin{equation}
\textrm{Tr}_{\mathcal{B}} \big(q^{L'_0-1}\big) = \frac1{\eta(\tau)^{24}}= q^{-1} + 24 + \cdots \ ,
\end{equation}
where $L'_0$ is the contribution of the oscillator modes to $L_0$.
\subsubsection*{The action of $M_{24}$ on $\mathcal{B}$: the first moonshine}

Let $g$ denote a symplectic automorphism of $K3$ of order $N$. Using the identification of the heterotic string as a NS5-brane wrapping $K3$, we can track the action of $g$ to the fields in the 1+1 dimensional CFT of the heterotic string. In particular, one can show that $g$ acts as a permutation of the 24 chiral bosons. Following the work of Mukai\cite{Mukai:1988}, it was shown in ref. \cite{Govindarajan:2009qt} that the possible conjugacy classes are those of $M_{24}$ that reduce to conjugacy classess of $M_{23}$ and order$(g)\leq8$ (and other conditions not stated here).  $M_{24}$ has precisely eight such conjugacy classes  that we list in Table \ref{cycleshapes}. As mentioned earlier, there exists symmetries of the $K3$ sigma model that are not symplectic automorphisms and do lead to other conjugacy classes that are not in the Table  \ref{cycleshapes}.
\begin{table}[hbt]
\centering
\newcommand\T{\rule{0pt}{2.6ex}}
\begin{tabular}{ccccccccc} \hline
$N$ & 1 & 2& 3 & 4 & 5 & 6 & 7 & 8\T \\[2pt] \hline 
$\rho$ & $1^{24}$ &  $1^82^8$ & $1^63^6$ & $1^42^24^4$ & $1^45^4$ & $1^22^23^26^2$ & $1^37^3$ & $1^2 2^14^18^2$ \T \\[3pt] \hline
Atlas & 1A & 2A & 3A & 4B & 5A & 6A & 7A & 8A\T \\[3pt] \hline
\end{tabular}
\caption{Table of cycle shapes (along with their  labels in the Atlas\cite{Atlasv3}) for symplectic automorphisms of $K3$.}\label{cycleshapes}
\end{table}

A straightforward  computation (not explicitly shown in ref. \cite{Govindarajan:2009qt} but given in ref. \cite[see Appendix A]{Govindarajan:2010fu}) leads to the following result:
\begin{equation}
\boxed{
\textrm{Tr}_{\mathcal{B}} \big(g \ q^{L'_0-1}\big) = \frac1{g_\rho(\tau)}= q^{-1} + a_1 + \cdots } \ ,
\end{equation}
where $\rho$ is the conjugacy class of $g$ and $g_\rho(\tau)$ is the $\eta$-product given by the map
\begin{equation}\label{cycleshapemap}
\quad \rho=1^{a_1}2^{a_2}\cdots N^{a_N} \longmapsto g_\rho(\tau)\equiv \prod_{j=1}^N\eta(j\tau)^{a_j}\ .
\end{equation}

\noindent \textbf{Remarks:}
\begin{enumerate}
\item There exist symmetries of the K3 sigma model that are not symplectic automorphisms. In particular, all other conjugacy classes of $M_{24}$ with at least four cycles, such as the cycle shape  $2^{12}$ (conjugacy class 2B),  arise this way\cite{Gaberdiel:2011fg}. 
The map Eq. \eqref{cycleshapemap} provides $\eta$-products for \textit{all} conjugacy classes of $M_{24}$.
\item The $\eta$-products that appear are multiplicative i.e., they are Hecke eigenforms. They also furnish a moonshine for the group $M_{24}$\cite{Dummit:1985,Mason:1985}. This is the first moonshine for $M_{24}$ that we will encounter -- this has been dubbed the \textit{additive} moonshine. As we will later see, this is the first of an infinite set of moonshines for $M_{24}$ that imply that $V^\natural$ is an $M_{24}$-module.
\item $\mathcal{B}$ is obviously an $S_{24}$-module and hence an $M_{24}$-module. However, $S_{24}$ is \textit{not} a symmetry of the full $D1-D5-KK-p$ system. For instance, the requirement that $g$ generate a symplectic automorphism needs us to choose special points in the moduli space of $K3$. This translates to a particular choice of the Narain lattice in the heterotic string -- as the Narain lattice does not form part of $\mathcal{B}$,  we see the  larger group i.e., $S_{24}$.
\item If $g$ has to generate a symplectic automorphism of $K3$, it forces us to restrict to conjugacy classes of $M_{24}$ that reduce to those of $M_{23}$ conjugacy. Other conjugacy classes also appear when one considers a pair of commuting symplectic automorphisms\cite{Govindarajan:2010fu}. These lead to a generalized moonshine (in the sense of Norton\cite[see appendix by Norton]{Mason:1987}) and will be discussed elsewhere.
\end{enumerate}

\subsection{The second-quantized elliptic genus}

The third contribution arises from the relative motion of the $D1$-branes in the worldvolume of $D5$-branes (in the type IIB frame). Consider the situation when one has $Q_1$ $D1$-branes wrapping $S^1$ and $Q_5$ D5-branes wrapping $K3\times S^1$. Recall that the magnetic charge of the system is $\tfrac12\mathbf{q}_m^2=(Q_1-Q_5)Q_5$. In particular, note that when $Q_5=Q_1=1$, the magnetic charge is vanishing. The other charges arise as momentum of the D1-brane along $S^1$ and $\widetilde{S}^1$. The BPS condition combined with level-matching in the associated CFT relates the dyonic charges to the $L_0$ and $J$ eigenvalues in the $RR$ sector.

In the $U(Q_5)$ supersymmetric gauge theory on coincident $D5$-branes, the $D1$-branes appear as instantons and the instanton number of the configuration is $(Q_1-Q_5)$ after taking into account the induced charge of a D5-brane wrapping $K3$\cite{Douglas:1995bn,Vafa:1995bm}.\footnote{When $Q_5=1$, one needs to turn on two-form flux to convert the $U(1)$ gauge theory to into a non-commutative  theory which admits instantons.}  The moduli space of this system is $4[(Q_1-Q_5)Q_5+1]$ dimensional and is identified with the symmetric product of $4[(Q_1-Q_5)Q_5+1]$ copies of $K3$\cite{Vafa:1995bm}. We denote the symmetric product of $m$ copies of $K3$  by $S^m(K3)\equiv K3^m/S_m$ and its smooth resolution by $K3^{[m]}$.

The dynamics of this relative motion is captured by the $\mathcal{N}=(4,4)$ superconformal sigma model with  target space $S^{m+1}(K3)$ where  $m=[(Q_1-Q_5)Q_5]$.  It was conjectured in ref. \cite{Dijkgraaf:1996it} and proved in ref. \cite{Dijkgraaf:1996xw} that the elliptic genus of this sigma model can be written in terms of the partition function of a second-quantized string theory on $K3\times S^1$. (See ref. \cite{Dijkgraaf:1998zd} for a review on the relation between second-quantized string theory and symmetric products.) Let us denote the ellliptic genus of a K\"ahler manifold, $M$,  by $\chi(M; \tau,z)$(see appendix \ref{ellipticgenus} for definitions) . Then, the result of ref. \cite{Dijkgraaf:1996xw} leads to a Borcherds product formula for the generating function of the elliptic genera of symmetric products of $K3$:
\begin{align}
\frac1{\mathcal{E}(K3; \mathbf{Z})}& := s^{-1}\times \Big( 1+\sum_{m=1}^\infty  s^{m} \ \chi(S^m(K3);\tau,z) \Big)\\
&=s^{-1}\times  \prod_{\substack{n\geq 0\\ m>0,\ell}} \frac1{(1-q^nr^\ell s^m)^{c(nm,\ell)}}\ , \label{Borcherdsproduct}
\end{align}
where $c(nm,\ell)$ are the Fourier-Jacobi coefficients of the elliptic genus of $K3$ which is a weight zero, index one Jacobi form that we denote by $\psi_{0,1}(\tau,z)$
\begin{equation}
\psi_{0,1}(\tau,z):= \chi(K3;\tau,z)= \sum_{n\geq0,\ell} c(n,\ell) \ q^n r^\ell \ .
\end{equation}
We refer to $\mathcal{E}(K3; \mathbf{Z})$ as the second-quantized elliptic genus of $K3$.   Since we associate  powers of  $s$ with the magnetic charge, we need to include an additional factor of $s^{-1}$ in our definition of the second-quantized elliptic genus. 

\noindent \textbf{Remark:} In the limit $z\rightarrow 0$, the elliptic genus, $\chi(M;\tau,z)$ reduces to the Euler characteristic of $M$ which we denote by $\chi[M]$. One sees that\footnote{For $m>1$, the manifold $S^m(K3)$  has orbifold singularities and by $ \chi[S^m(K3)]$, we mean the orbifold Euler characteristic of $S^m(K3)$.}
\begin{equation}
\lim_{z\rightarrow 0} \frac1{\mathcal{E}(K3; \mathbf{Z})} = \frac1s+ \sum_{m=1}^\infty \chi[S^m(K3)]\ s^{m-1} = \frac{1}{\eta(\sigma)^{24}}\ .
\end{equation}
The appearance of the $\eta$-product is easily understood by carrying out electric-magnetic duality and the \textit{magnetic} $\eta$-product gets mapped to the electric $\eta$-product. The condition that the ground state have charge $\mathbf{q}_m^2=-1$ then follows from level matching.

\subsection{Decomposing the module $\mathcal{S}$}

We decompose the module $\mathcal{S}$ using the magnetic charge as
\begin{equation}
\mathcal{S}:=\bigoplus_{m=-1}^\infty \mathcal{S}_m \ .
\end{equation}
The $m=-1$ term  is the \textit{ground state} with $Q_1=Q_5=0$ and has magnetic charge $\mathbf{q}_m^2=-1$ with other charges vanishing. Thus, $\mathcal{S}_{-1}=\mathbb{C}$. The sub-module $\mathcal{S}_0=\mathcal{H}(K3)$ is special and we shall call it $\mathcal{K}$ and the elliptic genus of $K3$ is given by
\begin{equation}
\psi_{0,1}(\tau,z) = \textrm{Tr}_{\mathcal{K}}\ \Big((-1)^Fq^{H_L} r^{J_L}\Big)\ .
\end{equation}
We will see that the modules $\mathcal{S}_m$ for $m\geq1$ can be constructed from $\mathcal{K}$\cite{Dijkgraaf:1996xw}. Recall $\mathcal{K}$ is the Hilbert space of a $D1$-brane  on $K3\times \widetilde{S}_1$ with winding number one. Denote by $\mathcal{K}_{(n)}$ the Hilbert space of a multiply wound $D1$-brane with winding number $n$ -- the oscillator modes of such a string have fractional moding $1/n$. Thus one has
\begin{equation}  
 \textrm{Tr}_{\mathcal{K}_{(n)}}\ \Big((-1)^Fq^{H_L} r^{J_L}\Big) = \psi_{0,1}\big(\tfrac{\tau}n,z\big)
\end{equation}

The work of DMVV\cite{Dijkgraaf:1996xw} uses the structure of pemutation orbifolds  and shows that the module $\mathcal{S}_{m-1}$ (for $m>1$)  can be decomposed into a direct sum of twisted sectors of the orbifold CFT. Recall that  twisted sectors of the $S_m$-orbifold are labelled by conjugacy classes of $S_m$. 
For instance, $\mathcal{S}_1$ has only two sectors -- the untwisted sector (with conjugacy class $1^2$) and a single twisted sector (with conjugacy class $2$). 
Thus, one has
\begin{equation}
\mathcal{S}_1 = \mathcal{H}_{1^2}^{\textrm{inv}} \oplus \mathcal{H}_{2}^{\textrm{inv}}\ ,
\end{equation}
where $\mathcal{H}_{1^2}=\mathcal{K}\otimes \mathcal{K}$ and $\mathcal{H}_2 = \mathcal{K}_{(2)}$ and the superscript `inv' indicates that we need to project onto $S_2$ invariant subspaces.
Thus the elliptic genus of $S^2(K3)$ can be written in terms of the elliptic genus of $K3$ and leads to the following \textit{doubling} formula.
\begin{align}
\chi(S^2(K3);\tau,z) &= \textrm{Tr}_{\mathcal{K}\otimes \mathcal{K}}\Big( \mathcal{P}q^{L_0-1} r^{J_L}(-1)^{F}\Big)+\textrm{Tr}_{\mathcal{K}_{(2)}} \Big(\mathcal{P}\ q^{L_0-1} r^{J_L}(-1)^{F}\Big)\ ,\nonumber
\\
&=\frac12\bigg[\big(\psi_{0,1}(\tau,z)\big)^2+\psi_{0,1}(2\tau,2z) \bigg]
+\frac12\bigg[\sum_{\ell=0}^1\psi_{0,1}(\tfrac{\tau+\ell}2,z)\bigg]\ , \label{doubling}
\end{align}
where $\mathcal{P}$ denotes the projection on to the $S^2$-invariant sector.

More generally, one has
\begin{equation}
\mathcal{S}_{m-1} = \bigoplus_{[h]} \mathcal{H}^{C_h}_{[h]}\ ,
\end{equation}
where $[h]$ denotes the conjugacy class of an element $h\in S_m$, $C_h$ is the centralizer group of $h$ and the superscript indicates that we project onto a $C_h$-invariant subspace of $ \mathcal{H}_{[h]}$. Thus, for a cycle shape $[h]=1^{a_1}2^{a_2}3^{a_3}\cdots $ (with $\sum_j j a_j=m$), the centralizer group is
$$
C_h = S_{a_1}  \times (S_{a_2}\rtimes \BZ_2^{a_2}) \times  (S_{a_3}\rtimes \BZ_3^{a_3})\times \cdots \ ,
$$
and $\mathcal{H}_{[h]}= \mathcal{K}^{\otimes a_1}\otimes \mathcal{K}_{(2)}^{\otimes a_2}\otimes \mathcal{K}_{(3)}^{\otimes a_3}\otimes \cdots$. Hence, one has
\begin{equation}
\mathcal{H}^{C_h}_{[h]} = S^{a_1}(\mathcal{K})\otimes S^{a_2}(\mathcal{K}^{\BZ_2}_{(2)})\otimes S^{a_3}(\mathcal{K}^{\BZ_3}_{(3)})\otimes \cdots \ .
\end{equation}
Thus, it is easy to see that the above structure leads to replication formulae i.e., all elliptic genera are expressible in terms of the elliptic genus of $K3$ i.e., $\psi_{0,1}(\tau,z)$.

DMVV show that the various replication formulae implied by the structure of the various $\mathcal{S}_m$ lead to the following remarkable formula:
\begin{equation}\label{untwistedformula}
\boxed{\Big( 1+\sum_{m=1}^\infty  s^{m} \ \chi(S^m(K3);\tau,z) \Big) = \exp\Big(\sum_{m=1}^\infty s^m\ V_m\cdot \psi_{0,1}(\tau,z)\Big)}\ ,
\end{equation}
where $V_m$ is the Hecke-like operator  that maps weak Jacobi forms of weight zero and index $1$ to weak Jacobi forms of weight zero and index $m$\cite{EichlerZagier}:\footnote{The Hecke-like operator is usually written as a sum over the coset $\Gamma_1\backslash S_m$ where $S_m$ is an $GL(2,\BZ)$ with $\det(S_m)=m$. In the following, we have written the Hecke-like operator for a particular parametrization of the coset.}
\begin{equation}
V_m\cdot \psi_{0,1}(\tau,z):=\sum_{\substack{ad=m\\b\textrm{ mod } d}} \frac1m\ \psi_{0,1}\big(\tfrac{a\tau+b}{d},az\big)\ .\label{untwistedHecke}
\end{equation}
It is easy to recover Eq. \eqref{doubling} by matching coefficients of $s^2$ on both sides of Eq. \eqref{untwistedformula}. On substituting the Fourier-Jacobi expansion of the elliptic genus of $K3$, we recover the Borcherds product formula given in Eq. \eqref{Borcherdsproduct}.

\subsection{The action of $M_{24}$ on $\mathcal{K}$: the second moonshine}

In CFT's on a torus on considers the following traces
\begin{equation}
\etabox{g}{h} = \Tr_{\mathcal{H}_h}\big (g\ \cdots \big)\ ,
\end{equation}
where $g$ and $h$ are symmetries of the CFT and $\mathcal{H}_h$ is a module twisted by $h$ (i.e., a module in the $h$-twisted sector in the orbifold by a group containing $h$). We will also denote the same object by $[g,h]$, on occasion, in a more compact notation.

There is growing  evidence that  $\mathcal{K}$ is an $M_{24}$-module\cite{Eguchi:2010ej,Cheng:2010pq,Gaberdiel:2010ch}. This is best seen by decomposing the elliptic genus of $K3$ in terms of characters of the $\mathcal{N}=4$ superconformal algebra\cite{Eguchi:2008gc,Eguchi:2009cq}. This is the \textit{second moonshine} for the Mathieu group $M_{24}$. As before, let $g$ be a symplectic automorphism of $K3$ of order $N$, then it has been shown that 
the $g$-twisted elliptic genus of $K3$ defined as follows
\begin{equation}
\boxed{
\psi_{0,1}^{[g,1]}(\tau,z):= \textrm{Tr}_{\mathcal{K}} \Big(g\ q^{L_0-1} r^{J_L}(-1)^{F}\Big)\ ,
}
\end{equation}
can be expressed in terms of characters of $M_{24}$ corresponding to the conjugacy class, $\rho$, of $g$. Several aspects have been numerically verified to fairly high powers by several authors\cite{Gaberdiel:2010ca,Eguchi:2010fg}. 
In this paper, we shall assume that $\mathcal{K}$ is indeed an $M_{24}$-module though this has not been proven to the best of our knowledge. With this assumption, we shall proceed to show that  $\mathcal{S}$ is  then an $M_{24}$ module in the next section. 

\section{The moonshine correspondence}

The generating function of $\tfrac14$-BPS states that are invariant under a symplectic automorphism $g$ (of order $N$) whose $M_{24}$ conjugacy class is $\rho=1^{a_2}2^{a_2}\cdots N^{a_N}$ is a Siegel modular form of weight $k$ and level $N$: $\Phi_{k}^{\rho}(\mathbf{Z})$ or more precisely $\Phi_{k}^{[g,1]}(\mathbf{Z})$  \cite{Govindarajan:2009qt}. As with $\Phi_{10}(\mathbf{Z})$, it can be written as product of three terms.\footnote{This Siegel modular form was called $\Phi^{(N,1)}_k(\mathbf{Z})$ in ref. \cite{Govindarajan:2010fu} and simply $\Phi_k(\mathbf{Z})$ in ref. \cite{Jatkar:2005bh}.}
\begin{equation}\label{quartergenfunction}
\frac{64}{\Phi_{k}^{[g,1]}(\mathbf{Z})} = \underset{\textrm{(i)}}{\underbrace{\bigg[\frac{4\ \eta(\tau)^6}{\theta_1(\tau,z)^2}\bigg]}} \times 
\underset{\textrm{(ii)}}{\underbrace{\bigg[\frac{16}{g_\rho(\tau)^{24}}\bigg]}}
\times \underset{\textrm{(iii)}}{\underbrace{\bigg[\frac1{\mathcal{E}^g\!\left(K3;\mathbf{Z}\right)}\bigg]}}\ ,
\end{equation}
Term (i)  remains unchanged while term (ii) becomes the multiplicative $\eta$-product as we discussed earlier. Term (iii) is the $g$-twisted second-quantized elliptic genus:
\begin{equation}
\frac1{\mathcal{E}^g(K3; \mathbf{Z})} := s^{-1}\times \Big( 1+\sum_{m=1}^\infty  s^{m} \ \chi^g(S^m(K3);\tau,z) \Big)\ ,
\end{equation}
where 
\begin{equation}
 \chi^g(S^{m+1}(K3);\tau,z) =  \textrm{Tr}_{\mathcal{S}_{m}}\Big(g\ (-1)^F q^{H_L} r^{J_L}\Big)\ ,
\end{equation}
is the $g$-twisted elliptic genus. There are symmetries of $K3$ that are not symplectic automorphisms that gives rise to other conjugacy classes of $M_{24}$. In these cases as well, we expect formula Eq. \eqref{quartergenfunction} to hold for the generating function of twisted $\tfrac14$-BPS states. However, it is not known whether they are modular forms.
The goal of this section is to provide evidence that $\mathcal{S}_m$ is an $M_{24}$-module graded by the $L_0$ and $J_L$ eigenvalues. Hence the above formula can interpreted as a trace over this module with $g$ taken to be an element of $M_{24}$. 

We have already seen that the modules $\mathcal{S}_m$ can be expressed in terms of $\mathcal{K}$. The `building blocks' of $\mathcal{S}_m$ are $S^m(\mathcal{K})$ and $\mathcal{K}_{(m)}$. If $g$ acts on $\mathcal{K}$, then it acts naturally as $\otimes^m g$ on the tensor product space
$\mathcal{K}^{\otimes m}$. This induces an action on $S^m(\mathcal{K})$.\footnote{On occasion, we shall be somewhat cavalier and use the same symbol $g$ to represent $\otimes^mg$ and its action can be inferred by the module on which it acts.} Given that $\mathcal{K}_{(m)}$ differs from $\mathcal{K}$ only in the $L_0$-grading and $g$ commutes with $L_0$, one may guess $g$ acts exactly as it did on $\mathcal{K}$. With this in mind, we shall consider $S_1$ and compute the $g$-twisted elliptic genus for $S^2(K3)$.
\begin{align}
\chi(S^2(K3);\tau,z) &= \textrm{Tr}_{S^2(\mathcal{K})}\Big(g\ q^{L_0-1} r^{J_L}(-1)^{F}\Big)+\textrm{Tr}_{\mathcal{K}^{\BZ_2}_{(2)}} \Big(g\ q^{L_0-1} r^{J_L}(-1)^{F}\Big)\ \hspace{3.5cm}\nonumber
\\
&=\frac12\bigg[\big(\psi^{[g,1]}_{0,1}(\tau,z)\big)^2+\psi^{[g^2,1]}_{0,1}(2\tau,2z) \big)
+\sum_{\ell=0}^1\psi^{[g,1]}_{0,1}(\tfrac{\tau+\ell}2,z)\bigg]\ . \label{newdoubling}
\end{align}
Notice the appearance of the $g^2$-twisted elliptic genus of $K3$ in the above formulae.
Thus we see that the $g$-twisted elliptic genus for $S^2(K3)$ satisfies a doubling formula as for the untwisted elliptic genus once we have included the action of $g$ on the various building blocks. We have experimentally verified that the above formulae holds when $g$ has order $2$ (conjugacy class 2A) and $3$ (conjugacy class 3A) -- this is possible as we already have explicit formulae for the second-quantized elliptic genus for  all conjugacy classes that appear in Table  \ref{cycleshapes}\cite{Govindarajan:2009qt}. Eq. \eqref{newdoubling} holds for all conjugacy classes of $M_{24}$ if we assume that $\mathcal{K}$ is a $M_{24}$-module.
We can thus anticipate  that replication formulae should be present for the $g$-twisted elliptic genus for all $S^m(K3)$. We propose that all such replication formulae can be written in terms of a \textit{twisted} Hecke-like operator, which we denote by $T_m$, that naturally generalizes the DMVV formula. 
\begin{equation}\label{twistedformula}
\boxed{\Big( 1+\sum_{m=1}^\infty  s^{m} \ \chi^g(S^m(K3);\tau,z) \Big) = \exp\Big(-\sum_{m=1}^\infty s^m\ T_m\cdot \psi^{[g,1]}_{0,1}(\tau,z)\Big)}\ ,
\end{equation}
where the twisted Hecke-like operator, $T_m$ defined below, generates a Jacobi form of weight $0$ and index $m$ when acting on a weak Jacobi form of weight $0$ and index $1$.
\begin{equation}\label{TwistedHeckeOperator}
\boxed{ T_m \cdot \psi_{0,1}^{[g,h]}(\tau,z) \equiv \frac1m \sum_{ad =m}\sum_{b=0}^{d-1}  \ \psi_{0,1}^{[g^ah^{-b},h^d]}\left(\tfrac{a\tau+b}{d},az\right)}\ . 
\end{equation}
The superscript $[g,h]$ indicates the boundary conditions. This extends the considerations of Tuite\cite{Tuite:2010}, who defined such an operator for modular forms, to Jacobi forms. The important ingredient is that the boundary conditions also change under the action of $GL(2,\BZ)$. One has
\begin{equation}
\etabox{g}h \stackrel{\gamma}{\longrightarrow} \etabox{g^ah^{-b}}{~~~g^ch^d}\ ,\quad \textrm{ for } \gamma=\left(\begin{smallmatrix} a & b \\ c & d \end{smallmatrix}\right)\in GL(2,\BZ)\ . 
\end{equation}

We have experimentally verified that the replication formulae  hold for twisted elliptic genera of $S^m(K3)$ for $m=2,3$ and for some choices of $g$ and $h$. In fact, we first  obtained these replication formulae experimentally before realizing that these are captured by the obvious generalization of Tuite's considerations to Jacobi forms. Such twisted Hecke-like operators have also appeared in the considerations of Ganter\cite{Ganter:2009,GanterThesis}. 

\subsection{An infinite number of moonshines}

We have seen that each of the spaces $\mathcal{S}_m$ are $M_{24}$-modules. Traces over them lead to Jacobi forms of weight zero and index $(m+1)$, one for each conjugacy class of $M_{24}$ appearing in Table 1. The first of these is the moonshine associated with the elliptic genus of $K3$. Of course, the Siegel modular form subsumes all these modular forms and hence the moonshine associated with $\Phi_{10}(\BZ)$ subsumes all these moonshines for $M_{24}$. It is important to note that on decomposing any of the Siegel modular forms into Jacobi forms and $\eta$-products, all of them are associated with the \textit{same} conjugacy class.

\section{Borcherds Product Formulae from moonshine}

We shall now show that Eq. \eqref{twistedformula} leads to a Borcherds product product formula for second-quantized twisted elliptic genus and hence for the corresponding Siegel modular form. These product formulae will be  shown to agree with known product formulae thus providing evidence for the conjectured form of the Hecke-like operator given in Eq. \eqref{TwistedHeckeOperator}. For other situations, in particular to all $g\in M_{24}$ that either have order $>8$ or have conjugacy classes that do not reduce conjugacy classes of $M_{23}$, we obtain Borcherds product formulae for such conjugally classes.

Let $g\in M_{24}$. The Fourier-Jacobi expansion of the weak Jacobi form $\psi_{0,1}^{[g^a,1]}(\tau,z)$ is
\begin{equation}
\psi_{0,1}^{[g^a,1]}(\tau,z) := \sum_{n=0}^\infty \sum_{\substack{\ell\in \BZ\\[2pt] 4n -\ell^2 \geq 0}} c^a(n,\ell)\ q^n r^\ell \ .
\end{equation}
When $g$ has order $N$,  $a$ is defined modulo $N$ since $\psi_{0,1}^{[g^a,1]}(\tau,z) =\psi_{0,1}^{[g^{b},1]}(\tau,z)$ for all $a\equiv b\textrm{ mod }N$. One has
\begin{align}
\sum_{b=0}^{d-1} \psi_{0,1}^{[g^a,1]}\Big(\tfrac{a\tau+b}d,a z\Big ) &=
\sum_{\wt{n}=0}^\infty \sum_{\substack{\ell\in \BZ\\[2pt] 4\wt{n} -\ell^2 \geq 0}} c^a(\wt{n},\ell)\ q^{a\wt{n}/d} r^{a\ell} \Big(\sum_{b=0}^{d-1} \beta^{b\wt{n}}\Big) \\
&= d\ \sum_{n=0}^\infty \sum_{\substack{\ell\in \BZ\\[2pt] 4nd -\ell^2 \geq 0}}c^{a}(dn,\ell)\ q^{an} r^{a\ell} \ ,
\end{align}
where $\beta:=\exp(2\pi i/d)$ is a $d$-th root of unity. In the second line, we have used the following result to set $\wt{n}=d n$
$$
\sum_{b=0}^{d-1} \beta^{b\wt{n}}= \left\{\begin{array}{cl} 0, & \wt{n}\neq 0\!\!\mod d\\[3pt]
d, & \wt{n}=0\!\!\mod d \end{array} \right.\ .
$$
Now consider the argument of the exponential in Eq. \eqref{twistedformula}
\begin{align}
\sum_{\widetilde{m}=1}^\infty s^{\widetilde{m}}\ T_{\wt{m}} \cdot \psi^{[g,1]}_{0,1}(\tau,z) 
&= \sum_{\widetilde{m}=1}^\infty \frac{s^{\widetilde{m}}}{\widetilde{m}} \sum_{\substack{ad=\widetilde{m}\\  a\in \BZ_{>0}}} \sum_{b=0}^{d-1} \psi_{0,1}^{[g^a,1]}\Big(\tfrac{a\tau+b}d,a z\Big )\nonumber  \\
&= \sum_{\widetilde{m}=1}^\infty  \sum_{\substack{ad=\widetilde{m}\\  a\in \BZ_{>0}}}  \sum_{n=0}^\infty \sum_{\substack{\ell\in \BZ\\[2pt] 4nd -\ell^2 \geq 0}}c^{a}(dn,\ell) q^{an} r^{a\ell}\frac{s^{ad}}{a}\nonumber  \\
&= \sum_{a=1}^\infty \sum_{m=1}^\infty \sum_{n=0}^\infty \sum_{\substack{\ell\in \BZ\\[2pt] 4nm -\ell^2 \geq 0}}c^{a}(nm,\ell)  \  \frac{q^{an} r^{a\ell}s^{am}}{a}\ ,\label{toproduct}
\end{align}
where we have reorganized the sum in the last row and relabeled $d$ as $m$.  Let the discrete Fourier transform of the Fourier-Jacobi coefficient be given by
\begin{equation}
c^a(n,\ell) = \sum_{\alpha=0}^{N-1} \omega^{\alpha a} c_\alpha(n,\ell)\ ,
\end{equation}
where $\omega=\exp(2\pi i/N)$ is an $N$-th root of unity. Using this transform, we can  carry out the summation over $a$ in the last line of Eq. \eqref{toproduct} to obtain
\begin{equation}
\sum_{\widetilde{m}=1}^\infty s^{\widetilde{m}} T_{\wt{m}} \cdot \psi^{[g,1]}_{0,1}(\tau,z) 
=-\sum_{\alpha=0}^{N-1} \sum_{m=1}^\infty \sum_{n=0}^\infty\!\! \sum_{\substack{\ell\in \BZ\\[2pt] 4nm -\ell^2 \geq 0}}   \log\Big(1-\omega^\alpha q^{n} r^{\ell}s^{m}\Big)^{c_{\alpha}(nm,\ell)}\ .
\end{equation}
Taking the exponential on both sides, we get
\begin{equation}
\exp\Big(-\sum_{\widetilde{m}=1}^\infty s^{\widetilde{m}} T_{\wt{m}} \cdot \psi^{[g,1]}_{0,1}(\tau,z) \Big)
=\prod_{\alpha=0}^{N-1} \prod_{m=1}^\infty \prod_{n=0}^\infty \!\!\prod_{\substack{\ell\in \BZ\\[2pt] 4nm -\ell^2 \geq 0}}   \Big(1-\omega^\alpha q^{n} r^{\ell}s^{m}\Big)^{c_{\alpha}(nm,\ell)}\ .
\end{equation}
Note that the product over $m$ runs from $1$ to $\infty$. This implies a Borcherds product formula for the associated Siegel modular form after multiplying by $s\ \phi^{[g,1]}_{k,1}(\tau,z)$. We obtain
\begin{equation}\label{productformulamain}
\boxed{
\Phi^{[g,1]}_k(\mathbf{Z}) = s\ \phi^{[g,1]}_{k,1}(\tau,z) \times \prod_{\alpha=0}^{N-1} \prod_{m=1}^\infty \prod_{n=0}^\infty \!\!\prod_{\substack{\ell\in \BZ\\[2pt] 4nm -\ell^2 \geq 0}}   \Big(1-\omega^\alpha q^{n} r^{\ell}s^{m}\Big)^{c_{\alpha}(nm,\ell)}
}\ .
\end{equation}
This formula is precisely the one obtained by David-Jatkar-Sen when $g$ generates a symplectic automorphism\cite[see Eq. (3.17)]{David:2006ud}. All one needs is to observe that 
$$\psi_{0,1}^{[g^a,1]}(\tau,z)=N F^{0,a}(\tau,z)\ ,$$ 
in their notation.  It can be shown that this formula  is equivalent to alternate versions of the formulae due to   Gritsenko-Nikulin\cite{GritsenkoNikulinII} and Aoki-Ibukiyama\cite{Aoki:2005} as well as an earlier formula also due to David-Jatkar-Sen\cite{David:2006ji}. In all other cases, we obtain a Borcherds product formula for the generating function of twisted $\tfrac14$-BPS states.

This concludes the proof that the twisted Hecke operator and the formula \eqref{twistedformula} for the twisted second-quantized elliptic genus holds. Further, with the assumption that  $\mathcal{K}$ is an $M_{24}$-module, we see that all modules $\mathcal{S}_m$ for $m>0$ are also $M_{24}$ modules. Thus, we obtain a self-consistent picture that the module $\mathcal{S}$ is an $M_{24}$-module.

\subsection{$M_{23}$ vs $M_{24}$}

The condition that any symplectic automorphism, $g$ must be contained in a $M_{23}$ subgroup of $M_{24}$ might seem to suggest that $V^\natural$ must be a $M_{23}$-module rather than a $M_{24}$ module. Recall that in the realization of $M_{24}$ as a permutation group,  $M_{23}$ is a subgroup of $M_{24}$ that preserves one element.
The key point is that the action of a symplectic automorphism depends only on its conjugacy class -- it is possible to find two different realizations of a symplectic automorphism, $g$, that  preserve distinct elements of $H^*(K3,\BZ)$ -- this is related to distinct symplectic structures.

There are two implications that arise from the claim that $V^\natural$ is a $M_{24}$-module. First, there are elements of $M_{23}$ with order $>8$ -- formally, one can insert such elements into the trace over $V^\natural$. Second, there exist elements of $M_{24}$ that are not in $M_{23}$. While we do not have an answer in full generality, we will comment on specific instances of these two possibilities below.
\begin{enumerate}
\item For instance, let us consider the case when $g$ has order $11$.  There is indeed a multiplicative eta product for the cycle shape $1^211^2$. The trace over the $V^\natural$ leads to the product formula given in Eq. \eqref{productformulamain} for a possible Siegel modular form of weight $0$. However, the naive formula for the additive lift does not work. Eguchi and Hikami have recently shown that there exists a modification due to the appearance of a new form for $\Gamma_0(11)$ and matches the terms appearing in the product formula\cite{Eguchi:2011aj}.
\item Since Jacobi forms have been constructed for all conjugacy classes of $M_{24}$\cite{Cheng:2010pq,Gaberdiel:2010ca,Eguchi:2010fg}, it appears that we must obtain a product formula for the corresponding elliptic genus and thence a product formula formula for a (potential) Siegel modular form for every conjugacy class of $M_{24}$ including those that do not reduce to conjugacy classes of $M_{24}$. One such conjugacy class is $2^{12}$ that has been recently shown to be a symmetry of the conformal field theory albeit not a symplectic one\cite{Gaberdiel:2011fg}. The Jacobi form as well as the eta product for the conjugacy class $2^{12}$ are both modular forms of $\Gamma_0(4)$. Should one expect a Siegel modular form of weight $4$ at level four? It has been argued in \cite{Govindarajan:2010fu} that the same cycle shape is associated with a  pair of symplectic automorphisms and that leads to a Siegel modular form of the paramodular group at level two. This appeared in the work of Clery and Gritsenko\cite{Gritsenko:2008}. We do not have a definitive answer on this case. So we will conclude this discussion with a couple of questions that we hope to address in the future. \textit{Should the twisted Hecke-like operator defined in Eq. \eqref{TwistedHeckeOperator} be modified for these conjugacy classes?} Indeed there is a modification that appears for the paramodular group in ref. \cite{Gritsenko:2008} and one needs to study if that is relevant for these cases.
\textit{Is there no factor that converts the second-quantized elliptic genus to a Siegel modular form for conjugacy classes of $M_{24}$ that do not reduce to $M_{23}$ conjugacy classes?} 
 It may be that the trace over the module $V^\natural$ is not a modular form -- recall that in all situations, the second-quantized elliptic genus needs to be multiplied by a  factor (called the Hodge anomaly by Gritsenko\cite{Gritsenko:1999}) to become a  modular form.  Note that an affirmative answer doesn't contradict the existence of $V^\natural$.  However, if the factor exists and we do obtain a modular form, it must arise as one of the modular forms constructed by Clery and Gritsenko\cite{Gritsenko:2008}. 
\end{enumerate}

\section{Conclusion}

In this paper, we have shown that the $D1-D5-KK-p$ system indeed provides  $M_{24}$-module. As a by-product of our investigation, we have implicitly obtained product formulae when the element $g$ has orders six and eight. These were already obtained using the additive (Saito-Kurokawa-Maa\ss) lift in ref. \cite{Govindarajan:2009qt}. It would be nice to see if there is Lie algebraic structure underlying these two modular forms of the kind when the order of $g$ was less than six\cite{Govindarajan:2008vi}. It is obvious that the replication formulae that we considered here also hold for the moonshine proposed by us in ref.  \cite{Govindarajan:2010cc} for the Mathieu group, $M_{12}$. This more or less follows from the observation in that paper that the twisted elliptic genus of $K3$   decomposes into to a sum of two terms along with a similar decomposition for the Siegel modular form. This implies that there exists an $M_{12}$-module, $\widehat{V}^\natural$ such that $V^\natural=\widehat{V}^\natural \otimes \widehat{V}^\natural$.

An obvious extension of our considerations is to consider the modular forms that arise in CHL orbifolds -- these were denoted by $\widetilde{\Phi}_k(\mathbf{Z})$ by Jatkar-Sen\cite{Jatkar:2005bh}. In the notation of this paper, these modular forms should be denoted by $\Phi^{[1,h]}_k(\mathbf{Z})$ where $h$ is an element of $M_{24}$.  As argued by us elsewhere\cite{Govindarajan:2010fu}, these should be considered in the context of generalized moonshine in the sense of Norton\cite[see appendix by Norton]{Mason:1987}. The analog of $V^\natural$ in this context should be a twisted version that we shall call $V^\natural_h$. Now consider insertions of another element $g$ of $M_{24}$ that commutes with  $h$ into the various trace. This should give rise  the modular forms that count twisted dyons in the CHL orbifold\cite{Govindarajan:2010fu}. The details of the generalized moonshine will be discussed elsewhere\cite{Govindarajan:2010cc}. These also provided realizations of conjugacy classes of $M_{24}$ that do not reduce to conjugacy classes of $M_{23}$. For instance, the class $\rho=2^{12}$ is associated with two commuting elements of $M_{24}$, each  of order two.  

\noindent \textbf{Acknowledgments:} We thank Karthik Inbasekar, Dileep  P. Jatkar and K. Gopala Krishna for comments on a draft of this paper. We also thank the organizers of the Conference on Modular Forms and Mock Modular Forms and their Applications in Arithmetic, Geometry and Physics held at ASICTP, Trieste during 14-18 March, 2011  as well as the organizers of the Workshop on Mathieu Moonshine (7-9 July 2011) at ETH, Z\"urich for an opportunity to present  this work.

\appendix

\section{The elliptic genus}\label{ellipticgenus}

Let $M$ denote a K\"ahler manifold of complex dimension $d$ and  $\mathcal{H}(M)$ denote the Hilbert space in the RR sector of the two-dimensional $\mathcal{N}=(2,2)$ Super Conformal Field Theory (SCFT) that arises from the supersymmetric nonlinear sigma model with target space $M$. The elliptic genus is given by\cite{Dijkgraaf:1996xw}
\begin{equation}
\chi(M; \tau,z) = \Tr_{\mathcal{H}(M)} \Big((-1)^{F} q^{H_L} \  r^{J_{L}}\Big)\ ,
\end{equation}
where $H_L = \big(L_0 -\tfrac{d}{8}\big)$, $J_{L}$ is the $U(1)$ R-charge for left-movers and $F=F_L+F_R$ is the total fermion number. When $M$ is a Calabi-Yau manifold as we shall assume henceforth,  the elliptic genus is a Jacobi form of weight $0$ and index $d/2$. Further, $\chi(M;\tau,0)$ is independent of $\tau$ and gives the Euler characteristic of $M$ which we denote by $\chi_M$. Let $g$ denote the generator of a  symmetry of the SCFT (of finite order) that commutes with the right-moving supersymmetry generators. One defines the $g$-twisted elliptic genus as follows: 
\begin{equation}
\chi^g(M; \tau,z) \equiv \Tr_{\mathcal{H}(M)} \Big(g \ (-1)^{F} q^{H_L}\ r^{J_L}\Big)\ .
\end{equation}
As in the untwisted case, the $z=0$  limit is $\tau$-independent and defines the twisted Euler characteristic, $\chi^g_M$. This obtains contributions from $g$-invariant harmonic forms when $g$ acts as  a permutation on $H^*(M,\BZ)$. In such cases, the $g$-twisted Euler characteristic is equal to the number of one-cycles in the conjugacy class of $g$ in the appropriate permutation group.

When $M$ is hyper-K\"ahler (with $d=2k$), then the supersymmetry is enhanced to $\mathcal{N}=(4, 4)$ and $J_{L}$ is identified with twice the Cartan generator of $SU(2)$ R-symmetry. Eguchi and Hikami\cite{Eguchi:2008gc,Eguchi:2009cq} have shown that the elliptic genus hyper-K\"ahler manifolds can be expanded in terms of massless and massive characters of the level $k$ $\mathcal{N}=4$ SCA. In  particular, one has (see \cite{Eguchi:2009ux} for precise definitions)
\begin{equation}
\chi(M; \tau,z) =  \chi_M\ \mathcal{C}_k(\tau,z) + \sum_{a=1}^k \Sigma^{(a)}(\tau) \ \mathcal{B}_k^{(a)}(\tau,z)\ ,
\end{equation}
where $\mathcal{C}_k(\tau,z)$ is a massless Ramond character (with isospin zero) and $\mathcal{B}^{(a)}_k(\tau,z)$ are massive Ramond characters. 
\begin{align}
\mathcal{C}^{(k)}(\tau,z) &= \textrm{ch}^{\tilde{R}}_{k,k/2,0} \nonumber \\
 &= \frac{\theta_1(\tau,z)^2}{\eta(\tau)^3} \frac{i}{\theta_1(\tau,2z)} 
\sum_{n\in\BZ} q^{(k+1)n^2} r^{2(k+1)n} \frac{1+q^n r}{1-q^nr}\ . \label{masslesschar}\\
\mathcal{B}_a^{(k)}(\tau,z) 
 &= \frac{\theta_1(\tau,z)^2}{\eta(\tau)^3} \frac{\vartheta_{k+1,a}(\tau,z)-\vartheta_{k+1,-a}(\tau,z)}{\vartheta_{2,1}(\tau,z)-\vartheta_{2,-1}(\tau,z) }\ ,\quad a=1,2,\ldots, k\ .\label{massivechar}
\end{align}
In particular, one has $\mathcal{C}_k(\tau,0)=1$ and 
$\mathcal{B}^{(a)}_k(\tau,0)=0$.

\subsection{A family of hyperK\"ahler manifolds}

Choose $M=K3$. The elliptic genus of $K3$ is a weight-zero index-one Jacobi form. One has 
\begin{equation}
\chi(K3;\tau,z) 
= 8 \sum_{i=2}^4 \left(\frac{\theta_i(\tau,z)}{\theta_i(\tau,0)}\right)^2 \ .
\end{equation}
Setting $z=0$, we see that $\chi_{K3}=24$ as expected. Further the elliptic genus of $K3$ can be expanded in terms of characters of the level-one $\mathcal{N}=4$ SCA.
\begin{equation}
\chi(K3;\tau,z) = 24\ \mathcal{C}_1(\tau,z) +\textrm{massive characters}\ .
\end{equation} 
An infinite family of hyperK\"ahler manifolds are obtained by considering symmetric products of $K3$\cite{Dijkgraaf:1996xw}. We will denote them by $S^n(K3)$ and are of complex dimension $d=2n$. The elliptic genus of $S^{n}(K3)$ is a  Jacobi form of weight-zero and index $n$ and can also be expanded in terms of characters of the level $n$ $\mathcal{N}=4$ SCA. In particular, one has
\begin{equation}
\chi(S^n(K3);\tau,z) = \chi_{S^n(K3)}\times  \mathcal{C}_n(\tau,z) +\textrm{massive characters}\ .
\end{equation} 
The space $S^n(K3)$ naturally arises as the moduli space of $n$ zero-brane probes or equivalently as the Hilbert scheme of $n$ points on a $K3$ surface\cite{Goettsche:1990}.

\bibliography{master}
\end{document}